\documentclass[namedreferences]{solarphysics}
\usepackage[optionalrh]{spr-sola-addons} 
\usepackage{enumerate}
\usepackage{graphicx}
\usepackage{color}
\usepackage{url}             

\begin{document}
\begin{article}
\begin{opening}

\title{Twin Null-Point-Associated Major Eruptive Three-Ribbon Flares with Unusual Microwave Spectra}

\author[corref,email={grechnev@iszf.irk.ru}]{\inits{V.V.}\fnm{V.V.}~\lnm{Grechnev}\orcid{0000-0001-5308-6336}}
\author[email={nata@iszf.irk.ru}]{\inits{N.S.}\fnm{N.S.}~\lnm{Meshalkina}}
\author[email={uralov@iszf.irk.ru}]{\inits{A.M.}\fnm{A.M.}~\lnm{Uralov}}
\author[email={kochanov@iszf.irk.ru}]{\inits{A.A.}\fnm{A.A.}~\lnm{Kochanov}\orcid{0000-0002-8079-1310}}
\author[email={svlesovoi@gmail.com}]{\inits{S.V.}\fnm{S.V.}~\lnm{Lesovoi}}
\author[email={ivan\underline{ }m@iszf.irk.ru}]{\inits{I.I.}\fnm{I.I.}~\lnm{Myshyakov}\orcid{0000-0002-8530-7030}}
\author[email={valentin\underline{ }kiselev@iszf.irk.ru}]{\inits{V.I.}\fnm{V.I.}~\lnm{Kiselev}}
\author[email={zhdanov@iszf.irk.ru}]{\inits{D.A.}\fnm{D.A.}~\lnm{Zhdanov}}
\author[email={altyntsev@iszf.irk.ru}]{\inits{A.T.}\fnm{A.T.}~\lnm{Altyntsev}}
\author[email={globa@iszf.irk.ru}]{\inits{M.V.}\fnm{M.V.}~\lnm{Globa}}

\address{Institute of Solar-Terrestrial Physics SB RAS,
                  Lermontov St.\ 126A, Irkutsk 664033, Russia}

 \runningauthor{V.V. Grechnev \textit{et al.}}
 \runningtitle{Twin Three-Ribbon Flares}

\date{Received ; accepted }

\begin{abstract}
On 23 July 2016 after 05:00\,UTC, the first 48-antenna stage of the
Siberian Radioheliograph detected two flares of M7.6 and M5.5 GOES
importance that occurred within half an hour in the same active
region. Their multi-instrument analysis reveals the following. The
microwave spectra were flattened at low frequencies and the spectrum
of the stronger burst had a lower turnover frequency. Each flare was
eruptive, emitted hard X-rays and $\gamma$-rays exceeding 800\,keV,
and had a rare three-ribbon configuration. An extended hard X-ray
source associated with a longest middle ribbon was observed in the
second flare. The unusual properties of the microwave spectra are
accounted for by a distributed multi-loop system in an asymmetric
magnetic configuration that our modeling confirms. Microwave images
did not to resolve compact configurations in these flares that may
also be revealed incompletely in hard X-ray images because of their
limited dynamic range. Being apparently simple and compact,
non-thermal sources corresponded to the structures observed in the
extreme ultraviolet. In the scenario proposed for two successive
three-ribbon eruptive flares in a configuration with a coronal-null
region, the first eruption causes a flare and facilitates the second
eruption that also results in a flare.
\end{abstract}


\end{opening}

\section{Introduction}
  \label{S-introduction}

Solar eruptions, flares, and similar weaker events draw energy from
coronal magnetic fields. Such events span over a vast range of
energy and time scales and spatial extent and manifest in various
associated phenomena. The way in which an eruption (flare) develops,
its manifestations and particularities depend on the magnetic
configuration that hosts the event and magnetic-field
transformations that occur in its course. Magnetic reconnection is
considered as the key process that governs solar eruptions and
flares.

Several properties of a two-ribbon flare and its development have
mainly been explained by the two-dimensional (2D) standard model
(CSHKP: \citealp{Carmichael1964, Sturrock1966, Hirayama1974,
KoppPneuman1976}). Reconnection in this model occurs in an X-point
of a vertical current sheet, which is formed due to the rise of a
prominence driven by a current instability \citep{Hirayama1974}. The
flare was implicitly presumed to be caused by the prominence
eruption, whose development was not considered. In modern models of
two-ribbon flares, the prominence is replaced by a magnetic
flux-rope. When the standard flare model is generalized to a 3D
situation, new particularities appear in magnetic reconnection and
the shapes and location of paired flare ribbons that are absent in
2D models (see, e.g., \citealp{Aulanier2012, Janvier2013} and
references therein).

Two conditions are necessary for the development of a large-scale
current instability that leads to the flux-rope expansion. These are
i)~formation in the corona of a flux rope, where the twist [$N$] of
magnetic-field lines certainly exceeds unity ($N > 1$), and ii)~its
rise to a height, starting from which the expansion becomes
continuous. The second condition is met, if the component of the
external magnetic field transversal to the flux-rope axis falls off
with height fast enough.

The way widely used for the formation of a magnetic flux-rope and
initiation of a two-ribbon flare invokes as boundary conditions
different types of photospheric plasma motions such as shear,
converging, or rotational motions (e.g.
\citealp{InhesterBirnHesse1992, Longcope2007}). A shorter way is
also possible. In the dual-filament model (\citealp{Uralov2002,
Grechnev2006erupt}; see also \citealp{HansenTripathiBellan2004}),
low-corona reconnection between two or more filament sections and
between their threads (barbs) increases the length and height of the
combined filament. This increases its dipole momentum and the total
twist up to $N > 1$, launching the standard-model reconnection.

Relatively recently, 3D coronal configurations with a null-point
topology (NPT) have been identified above photospheric magnetic
islands surrounded by opposite-polarity environment (e.g.
\citealp{Filippov1999, Filippov2009, Masson2009, Meshalkina2009,
Pariat2009, Pariat2010, Reid2012, WangLiu2012}). Such configurations
are ubiquitous, being responsible for a variety of phenomena from
tiny polar X-ray jets up to major flares. A flare that occurs in an
NPT-configuration produces a circular ribbon with a brightening in
the center. An implicated remote compact site is sometimes
detectable.

According to \cite{Filippov2009} and \cite{Meshalkina2009}, a
reconnection event in an NPT-configuration is generally initiated
and governed by the eruption of a small flux-rope-like structure
that occurs inside an inverted-funnel-like separatrix surface.
Reconnection behind the erupting structure occurs in the same way as
in a usual two-ribbon flare. When the eruption passes through the
null-point region, its magnetic structure becomes partly or entirely
destroyed. A jet is ejected in the latter case (see also
\citealp{Sterling2016} who came to similar conclusions). This
sequence of events is supported by a half-minute delay of the hard
X-ray burst relative to the acceleration of the erupting structure
measured by \cite{Grechnev2011_I} in the event that
\cite{Meshalkina2009} considered, where a bright ring-like structure
and broad jet were observed in the extreme ultraviolet. Conversely,
the numerical magnetohydrodynamic (MHD) simulations of
\cite{Pariat2009, Pariat2010} and \cite{Masson2009, Masson2017} do
not require small flux-rope eruptions as significant components of
their NPT-associated models of flares and jets (see also a review by
\citealp{Raouafi2016}).

Larger-scale phenomena probably caused by reconnection between
magnetic structures of erupting filaments and static coronal
environment have also been sometimes observed. They are manifested
in bifurcation of an erupting structure and dispersal of the erupted
material over a large surface far from the eruption region
\citep{Slemzin2004, Grechnev2005, Grechnev2014_I, Uralov2014}. The
dispersed low-temperature material released by reconnection from the
filament body can screen large areas on the Sun and cause an
extensive darkening in 304\,\AA\ and depression of the total
microwave flux \citep{Grechnev2008blast, Grechnev2011anomal,
Grechnev2013neg}. Reconnection in such events is forced by the
expansion of an erupting structure that encounters a topological
obstacle in its path in the corona. A similar conclusion was drawn
by \cite{vanDriel2014} for the spectacular SOL2011-06-07 event.

While MHD simulations demonstrate the possibility of reconnection in
an NPT-configuration without any eruption, observations indicate
that such phenomena are caused by interactions of erupting
structures with static coronal magnetic fields. In this respect,
flares in NPT-configurations do not seem to be considerably
different from usual eruptive two-ribbon flares, where flare
processes are caused by eruptions that basically corresponds to the
scenario of \cite{Hirayama1974}. This relation is supported by the
correspondence between the kinematics of an erupting structure and
flare emissions, which were delayed by up to two minutes that was
found in studies of several eruptive events of different types and
importance \citep{Grechnev2011_I, Grechnev2013_20061213,
Grechnev2015, Grechnev2016, Grechnev2019}. On the other hand, in the
absence of eruptions, coronal null points are steady and do not
reveal themselves in any way (e.g. \citealp{Filippov1999,
Grechnev2014_I}) that does not favor the scenario, in which an event
starts from the null-point reconnection itself.

The morphology was still more outstanding in the events presented by
\cite{Wang2014}. The authors addressed a pair of successive flares
(M1.9 and C9.2) that occurred on 6 July 2012 within half an hour and
both exhibited unusual three-ribbon configuration. The flares were
accompanied by surges and jets. The authors conjectured that the
events were caused by reconnection along the coronal null line in
the fan--spine magnetic topology. \cite{Bamba2017} presented a
three-ribbon flare that occurred on 25 October 2014.

We consider a pair of three-ribbon flares that also occurred within
half an hour in the same active region on 23 July 2016. The flares
(M7.6 and M5.5) were stronger than those addressed by
\cite{Wang2014}, both of them were eruptive and produced conspicuous
emissions in microwaves, hard X-rays, and $\gamma$-rays. The clear
presence of filament eruptions in both events removes the question
of their possible involvement. The large size of the flares favors
the analysis of various manifestations that were barely detectable
in weaker events. Taking advantage of these unprecedented
observations, we address the properties of the two eruptive flares,
reveal the magnetic configuration where they occurred, and pursue
understanding their common scenario.

Section~\ref{S-overview} briefly overviews the twin events and
highlights their particularities.
Section~\ref{S-configuration_scenario} addresses the coronal
configuration, where the events occurred, and infers their scenario.
Section~\ref{S-acc_electrons} addresses manifestations of
accelerated electrons in microwave and hard X-ray images and
considers the spectra in the two spectral domains.
Section~\ref{S-conclusion} summarizes the results. The
\url{AIA335.mpg}, \url{pot_Br_mod_B.mpg}, and
\url{nlff_Br_mod_B.mpg} movies in the Electronic Supplementary
Material illustrate the twin events and particularities of the
coronal magnetic configuration.

\section{Overview of the Twin Events}
 \label{S-overview}

\subsection{Microwave and Hard X-Ray Non-Imaging Data}
 \label{S-microwaves_HXR}

The Siberian Radioheliograph (SRH: \citealp{Lesovoi2014,
Lesovoi2017}) commenced test-mode observations early in 2016. In
summer 2016, the SRH started to observe the Sun routinely at five
frequencies of 4.5, 5.2, 6.0, 6.8, and 7.5\,GHz along with the
ongoing adjustment of its systems. The temporal interval to scan the
five frequencies was 8.4 seconds at that time. The longest SRH
baseline of 107.4\,m determines its spatial resolution of order
$100^{\prime \prime }$ (depending inversely on frequency) that
allows locating a microwave source on the Sun but is insufficient to
reveal its shape and structure. Real-time beacon SRH data at a set
of the operating frequencies with an update every minute are
accessible online at the SRH Web site \url{badary.iszf.irk.ru/}.

On 23 July 2016, the SRH recorded strong microwave bursts associated
with two major flares that occurred within half an hour at N05\,W73
in active region, whose parts were numbered 12565 and 12567. The
GOES importance was M7.6 for the first flare and M5.5 for the second
flare. The Badary Broadband Microwave Spectropolarimeters (BBMS:
\citealp{Zhdanov2011, Zhdanov2015, Kashapova2013}) that are
installed near the SRH measured the total flux up to 400\,sfu in the
first flare and up to 800\,sfu in the second flare.

\begin{figure} 
  \centerline{\includegraphics[width=0.8\textwidth]
   {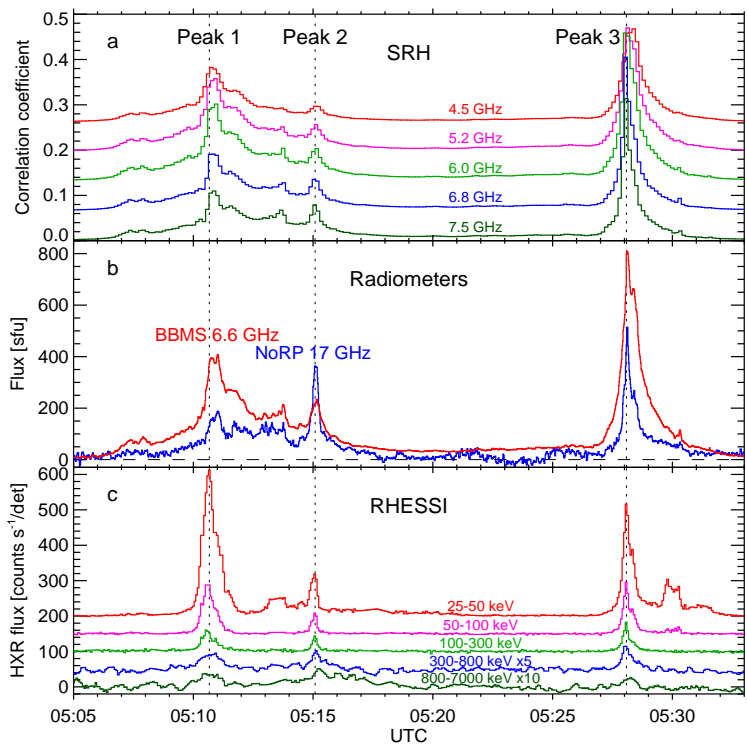}
  }
  \caption{Temporal profiles of the two flare bursts.
(\textbf{a})~Correlation plots recorded by the SRH.
(\textbf{b})~Microwave bursts recorded by total-flux radiometers at
6.6\,GHz (BBMS) and 17\,GHz (NoRP). (\textbf{c})~Hard X-ray and
$\gamma$-ray bursts observed by RHESSI. Two highest-energy count
rates are smoothed and magnified by factors of 5 (300\,--\,800\,keV)
and 10 (800\,--\,7000\,keV). The background levels on panels
\textbf{a} and \textbf{c} are shifted to show the bursts better. The
vertical dotted lines mark three main peaks that are detectable in
all ranges.}
  \label{F-time_prof}
  \end{figure}

Figure~\ref{F-time_prof} shows the temporal profiles recorded during
the two flares in microwaves and hard X-rays (HXR).
Figure~\ref{F-time_prof}a presents the so-called correlation plots
produced by the SRH that are computed without synthesizing the
images as the sum of cross-correlations between all antenna pairs
\citep{LesovoiKobets2017}. The correlation plots represent a proxy
of total-flux variations, responding to the changes in both the
brightness and structure of microwave sources. To get indications at
total-flux spectra of the sources that were unresolved in these
observations, each $i$-th correlation plot obtained at a frequency
$\nu_i$ was multiplied by $\left(\nu_i/\overline{\nu}\right)^2$,
where $\overline{\nu}$ is the average frequency of the SRH observing
range.

Three main peaks numbered 1\,--\,3 are distinct. Comparison of the
peaks at different SRH frequencies indicates that the turnover
frequencies (peak frequencies of the spectra) for peaks 1 and 3 were
within the SRH range or nearby and higher for peak~2. These
indications are confirmed in Figure~\ref{F-time_prof}b that presents
total-flux variations recorded by the BBMS at 6.6\,GHz and by the
Nobeyama Radio Polarimeters (NoRP: \citealp{Torii1979,
Nakajima1985}) at 17\,GHz. The peaks had different spectra indeed:
while peaks 1 and 3 were stronger at 6.6\,GHz, peak 2 was stronger
at 17\,GHz.

Figure~\ref{F-time_prof}c presents HXR and $\gamma$-ray bursts
recorded by the Reuven Ramaty High-Energy Solar Spectroscopic Imager
(RHESSI: \citealp{Lin2002}) in five energy bands. To reduce the
variable background, whose importance increases with increasing
energy, its variations in the preceding and following orbits were
fitted for each energy band and subtracted. To enhance the
signal-to-noise ratio in the two highest-energy bands plotted, the
count rates were smoothed with a boxcar average over three neighbors
for the 300\,--\,800\,keV band and over five neighbors for the
800\,--\,7000\,keV band. They are magnified in
Figure~\ref{F-time_prof}c. The three main peaks became detectable at
energies exceeding 800\,keV.

It is possible to judge qualitatively about the photon HXR spectra
by comparing visually the heights of the peaks in a high-energy band
(e.g. 300\,--\,800\,keV) and in the 25\,--\,50\,keV band. Peak 1 had
the softest photon spectrum, peak 2 had the hardest spectrum, and
the spectrum of peak 3 was in between of them. The same relation is
expected for high-frequency parts of the microwave spectra of the
three peaks \citep{DulkMarsh1982, White2011}.

\begin{figure} 
  \centerline{\includegraphics[width=0.65\textwidth]
   {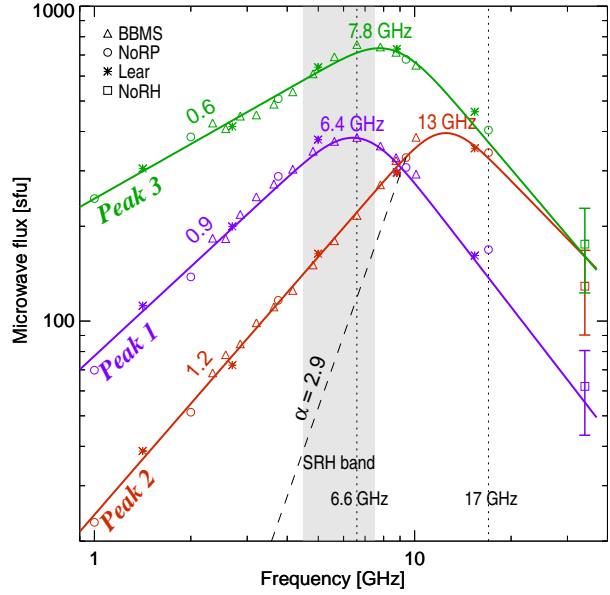}
  }
  \caption{Microwave spectra of the three peaks denoted in
Figure~\ref{F-time_prof} composed from measurements at different
total-flux radiometers and NoRH (symbols). The thick color curves
represent their simplest fit, from which the peak frequencies
indicated near the spectral peaks were estimated. The slopes of the
high-frequency branches have considerable uncertainties. The slanted
dashed line demonstrates the theoretical slope $\alpha = 2.9$
expected at low frequencies for a single gyrosynchrotron source if
it were responsible for peak~2. The shading denotes the SRH
observing frequency range. The vertical dotted lines denote the
frequencies of 6.6 and 17\,GHz, at which the time profiles presented
in Figure~\ref{F-time_prof}b were measured.}
  \label{F-mw_spectra}
  \end{figure}

Figure~\ref{F-mw_spectra} present total-flux microwave spectra of
integrated over eight seconds around each peak. We used one-second
NoRP data, one-second data obtained at the Learmonth station of the
US Air Force Radio Solar Telescope Network (RSTN:
\citealp{Guidice1979, Guidice1981}), and BBMS data with a temporal
sampling of 1.6 seconds. The whole data set had problems. The BBMS
consists of two instruments, each of which employs its own spectrum
analyzer. The 4\,--\,8\,GHz spectropolarimeter operated in the
high-sensitivity mode so that its data were saturated. A few
channels of the 2\,--\,24\,GHz spectropolarimeter were out of
operation. The triangles in Figure~\ref{F-mw_spectra} in the
2.34\,--\,10.1\,GHz range represent the data from the BBMS channels
that appear to be reliable.

To find the slope of the descending part of the microwave spectrum,
high-frequency measurements are important. However, the total-flux
variations shown by NoRP at 35\,GHz are strongly dissimilar to the
temporal profiles recorded at lower frequencies. This indicates
improper operation of the 35\,GHz instrument. We had to use instead
the images produced by the Nobeyama Radioheliograph (NoRH:
\citealp{Nakajima1994}) at 34\,GHz. NoRH suffered in this period
from hardware problems. We synthesized NoRH images at 34\,GHz around
each of the three peaks with intervals and integration times of one
second and five seconds. The images in each set were coaligned with
each other. To calibrate the images in brightness temperatures, the
regions of the solar disk and the sky were analyzed separately, and
the most-frequent pixel values were referred to $10^4$\,K and zero,
respectively \citep{Kochanov2013}. The one-second image set was
subjected to the median smoothing along each pixel with widths of
three and five. Then, the total flux was computed over the flaring
region from all image sets. Comparison of the results with each
other indicates the calibration stability of the reduced-quality
NoRH 34\,GHz images of about $\pm 30\%$. The squares in
Figure~\ref{F-mw_spectra} represent probable fluxes at 34\,GHz with
the $\pm 30\%$ uncertainties shown by the bars.

The ascending and descending branches of the microwave spectra were
analyzed separately using the linear fit in the log--log scale.
Because of the insufficient measurement accuracy at 34\,GHz, the
slopes of the high-frequency branches have increased uncertainties,
especially for peak~2. The two branches were connected with the
antiderivative (indefinite integral) of the error function. The peak
frequencies estimated from the fitting curves are indicated at the
curves along with low-frequency slopes [$\alpha$] found from the
linear fit. The positions of the estimated peak frequencies relative
to the SRH frequency range (the shading) are consistent with the
assessments that were made from the correlation plots in
Figure~\ref{F-time_prof}a, while the ongoing adjustment of the SRH
hardware systems disfavored accurate calculations of the flux
spectrum from the images at that time.

The low-frequency branches of the spectra are flattened with respect
to the theoretical slope $\alpha = 2.9$ expected for a simple single
gyrosynchrotron source. This slope is shown by the dashed line for
peak~2. A possible cause of the low-frequency spectral flattening
may be asymmetry of the flaring magnetic configuration; the
magnetic-flux balance requires a larger area of a weaker-field side
than of the conjugate region that elevates the low-frequency
spectral part and shifts the peak frequency to the left
\citep{Grechnev2017}. The change from peak 2 to peak 3 seems to be
especially challenging, because no change in most of the parameters
that govern gyrosynchrotron emission displaces the top of its
spectrum up and left \citep{Stahli1989}. These circumstances
indicate that the flare configuration was asymmetric and relatively
complex.

\subsection{Two Successive Eruptions}
 \label{S-two_eruptions}

The paired events comprised two sequential eruptions that were
observed in all extreme-ultraviolet (EUV) channels of the
Atmospheric Imaging Assembly (AIA) on board the Solar Dynamics
Observatory (SDO: \citealp{Lemen2012}). The two eruptions are
demonstrated by the \url{2016-07-23_AIA335.mpg} Electronic
Supplementary Material. The movie was composed from the images
obtained in the low-sensitivity 335\,\AA\ AIA channel, which did not
suffer from saturation.

\begin{figure} 
  \centerline{\includegraphics[width=\textwidth]
   {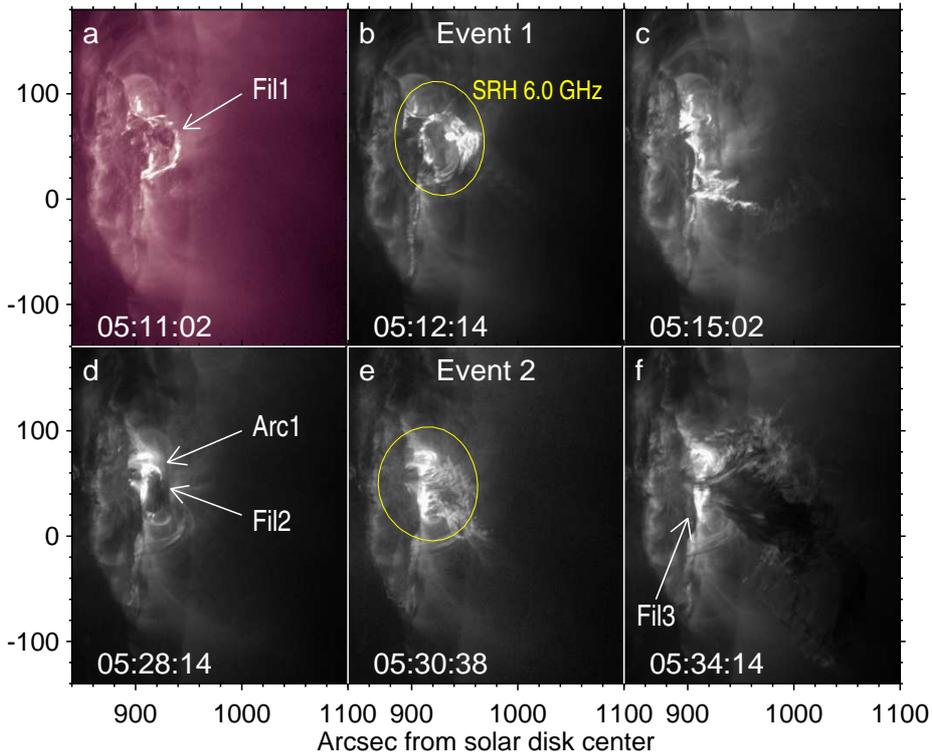}
  }
  \caption{Two eruptive events observed within half an hour
by SDO/AIA in 211\,\AA. The yellow contours on panels \textbf{b} and
\textbf{e} represent the SRH 6.0\,GHz images at half-height levels.
The arrows indicate eruptive filaments Fil1 and Fil2. Arcade Arc1
formed in the first event is visible. Filament Fil3 did not erupt
and partly blocked in EUV the second, southern arcade.}
  \label{F-two_events}
  \end{figure}

Figure~\ref{F-two_events} presents a few episodes of the two
eruptive events. Event~1 started from the eruption of the first
filament, Fil1, in Figure~\ref{F-two_events}a that corresponds to
the time of peak~1 in Figure~\ref{F-time_prof}. The filament
brightened up, inflated (Figure~\ref{F-two_events}b), and expanded
further. The ragged appearance of the erupting filament in
Figure~\ref{F-two_events}c indicates that its structure became
partly damaged. Arcade Arc1 in Figure~\ref{F-two_events}d appeared
in the first event, whose GOES importance reached a level of M7.6.
The yellow contour in Figure~\ref{F-two_events}b presents the
unresolved microwave source observed by the SRH at 6.0\,GHz at a
half-height level that is close to the SRH beam.

The second event associated with the eruption of another filament,
Fil2 (Figures \ref{F-two_events}d\,--\,\ref{F-two_events}f), was
mainly similar to the first event. Figure~\ref{F-two_events}d
corresponds to the time of peak~3 in Figure~\ref{F-time_prof}. The
structure of the second filament was also damaged; dark filament
material dispersed and partly returned back to the solar surface
(Figure~\ref{F-two_events}f). The third filament, Fil3, denoted in
Figure~\ref{F-two_events}f exhibited motions, but did not erupt.
This filament partly hid the southern flare arcade in EUV. The GOES
importance of the second event was M5.5. The yellow contour in
Figure~\ref{F-two_events}e presents the 6.0\,GHz SRH image (similar
to Figure~\ref{F-two_events}b).

The two-fold character of the whole event manifested in associated
phenomena. Each of the two eruptions produced an EUV wave
\citep{Chandra2018}. The online CME catalog
(\url{cdaw.gsfc.nasa.gov/CME_list/}: \citealp{Yashiro2004}) presents
an associated CME, whose faster northern part was followed by a
slower southern part that was launched apparently later.

\subsection{Three-Ribbon Flare Configuration}
 \label{S-three_ribbon}

\begin{figure} 
  \centerline{\includegraphics[width=\textwidth]
   {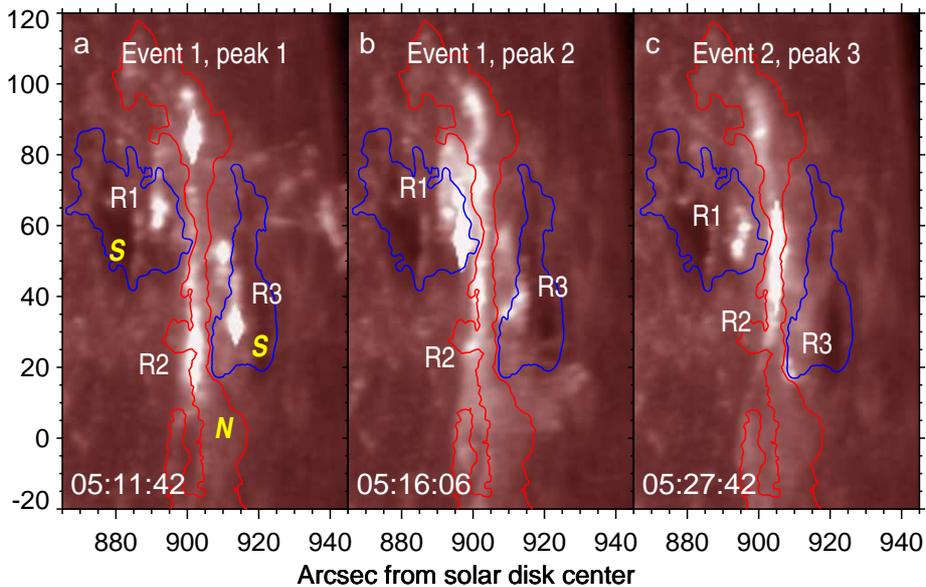}
  }
  \caption{Three-ribbon configurations observed by SDO/AIA
in 1700\,\AA\ in the two flares near the three main peaks. The blue
contours outline the S-polarity radial magnetic component, and the
red contour outlines the N-polarity region. The contour levels are
$[-240, +120]$\,G.}
  \label{F-three_ribbons}
  \end{figure}

A particularity of the two corresponding flares was their
three-ribbon configuration. High brightness of the flare ribbons
caused strong overexposure distortions in the 1600\,\AA\ SDO/AIA
channel. The 1700\,\AA\ AIA images were also saturated, but their
quality was better. Figure~\ref{F-three_ribbons} presents the flare
ribbons observed by AIA in 1700\,\AA\ in the two flares near the
three main peaks. The contours overlaid on the images represent the
polarities of the radial magnetic component that was computed from
vector magnetograms produced on 23 July by the Helioseismic and
Magnetic Imager (HMI: \citealp{Scherrer2012}) on board SDO.

Each of the three panels in Figure~\ref{F-three_ribbons} shows a
long N-polarity middle ribbon R2 and two considerably shorter
ribbons R1 and R3 in S-polarity regions on both sides of the middle
ribbon. The three-ribbon configuration corresponds to the
S\,--\,N\,--\,S structure of ribbons R1\,--\,R2\,--\,R3. The
magnetic fields within the contours were stronger in the S-polarity
regions than in the N-polarity region by a factor of about 2.3 on
the average. The shorter lengths of the S-polarity ribbons are
determined by the magnetic-flux balance at the conjugate regions.
This circumstance confirms the indications suggested by the shapes
of the microwave spectra that were outlined in
Section~\ref{S-microwaves_HXR}.

\section{Coronal Magnetic Configuration and Scenario of the Events}
 \label{S-configuration_scenario}

\subsection{Magnetic Configuration}
  \label{S-configuration}

\begin{figure} 
  \centerline{\includegraphics[width=\textwidth]
   {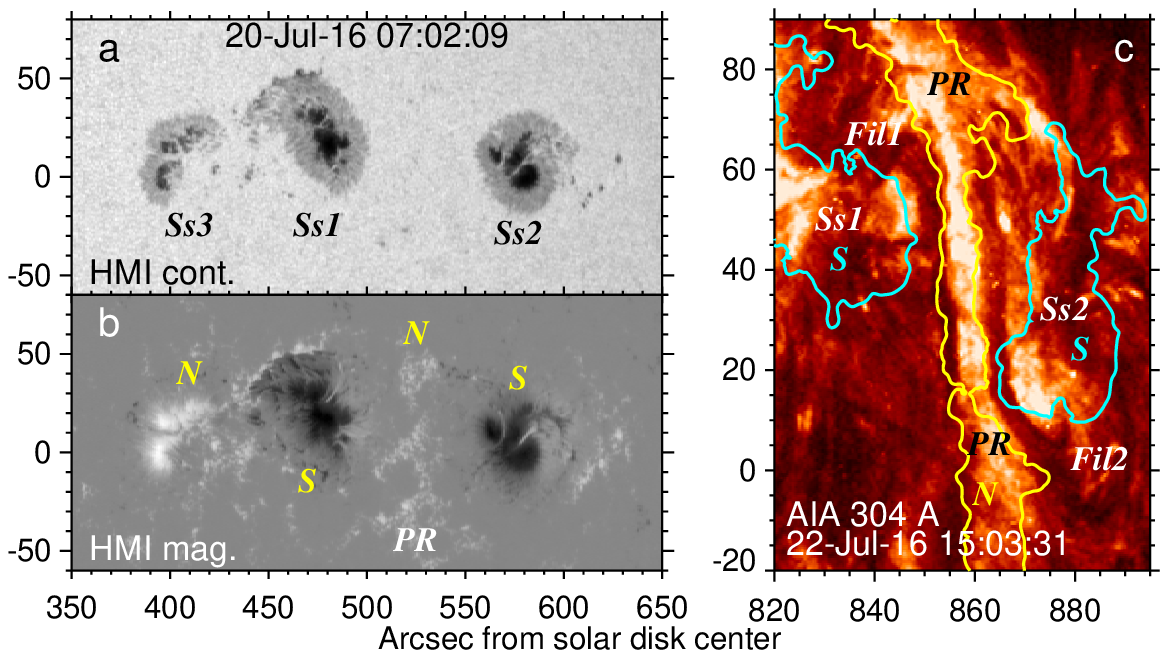}}
  \vspace{0.2cm}
  \centerline{\includegraphics[width=0.6\textwidth]
   {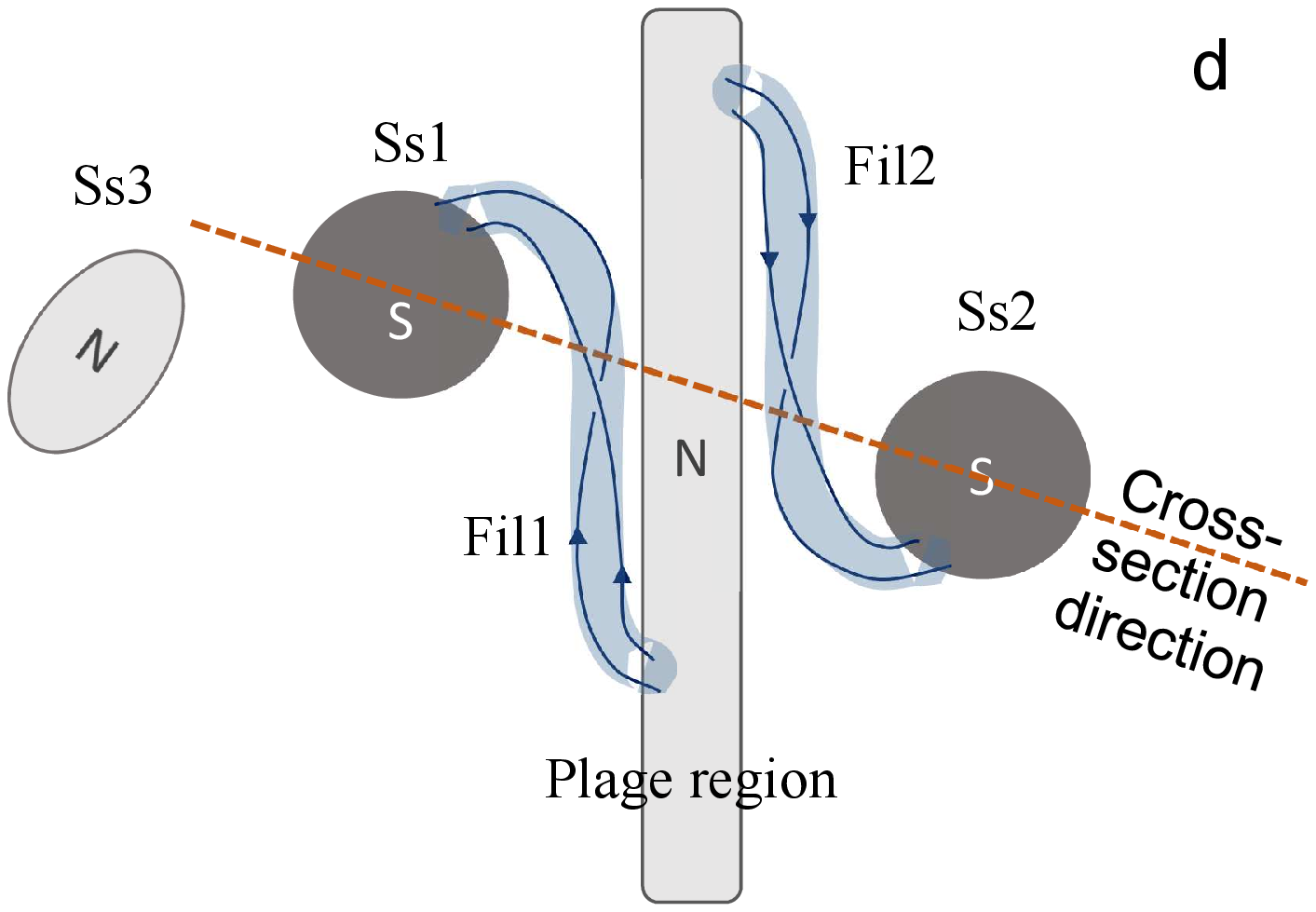}
  }
  \caption{Pre-flare configuration. Intensitygram (\textbf{a})
and line-of-sight magnetogram (\textbf{b}) of the active region
observed by SDO/HMI on 20 July, three days before the events. The
yellow labels N and S denote magnetic polarities of sunspots Ss1,
Ss2, and Ss3, and plage region PR. (\textbf{c})~SDO/AIA 304\,\AA\
image observed on 22 July presents two filaments Fil1 and Fil2
located between the S-polarity sunspots Ss1 and Ss2 and N-polarity
plage region PR. (\textbf{d})~A scheme of the magnetic configuration
inferred from observations. The slanted brown-dashed line denotes
the direction of the cross-section that is presented in
Figure~\ref{F-scenario}.}
  \label{F-mag_config}
  \end{figure}

The 23 July twin events were located not far from the limb, where
considerable projection shrinkage complicates the understanding of
the magnetic configuration. We firstly consider observations of the
active region, which hosted the events, that were made a few days
before. The active region observed by SDO/HMI on 20 July is shown in
Figures \ref{F-mag_config}a (intensitygram) and \ref{F-mag_config}b
(line-of-sight magnetogram). Two S-polarity sunspots Ss1 and Ss2
were separated by an extended N-polarity plage region PR. No
conspicuous flare manifestations were visible near the eastern
N-polarity sunspot Ss3. The magnetic field in the plage region
strengthened from 20 to 23 July.

Figure~\ref{F-mag_config}c presents a 304\,\AA\ AIA image observed
14 hours before the events along with contours of the radial
magnetic component computed from a simultaneous HMI vector
magnetogram. Two filaments are visible in 304\,\AA. Filament Fil1
was rooted by the northern end in the S-polarity environment of
sunspot Ss1 and mostly located above the neutral line between the
Ss1 environment and the plage region PR. Filament Fil2 was rooted by
the southern end in the S-polarity environment of sunspot Ss2 and
mostly located above the neutral line between the Ss2 environment
and PR. The opposite ends of both filaments must be rooted in an
N-polarity region, i.e. in the plage region PR. These circumstances
are shown in the scheme of the magnetic configuration presented in
Figure~\ref{F-mag_config}d.

\begin{figure} 
  \centerline{\includegraphics[width=0.52\textwidth]
   {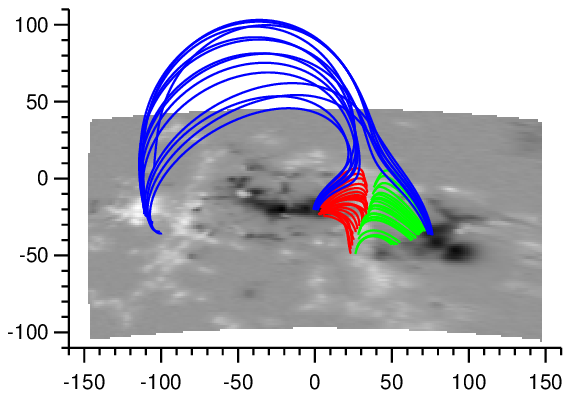}
  \hspace*{-0.02\textwidth}
          \includegraphics[width=0.52\textwidth]
   {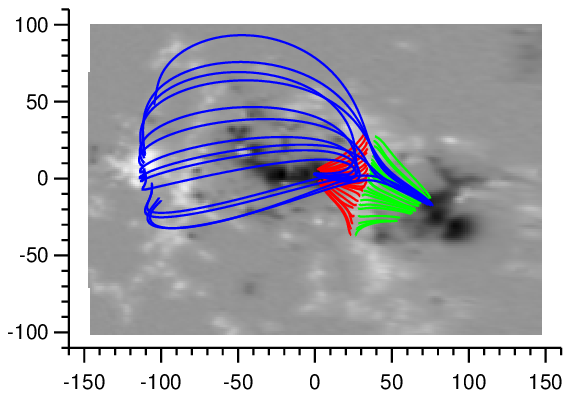}
  }
     \vspace{-0.316\textwidth}
     \centerline{\large \bf
      \hspace{0.375 \textwidth}  \color{black}{a}
      \hspace{0.445\textwidth}   \color{black}{b}
         \hfill}
     \vspace{0.292\textwidth}
  \caption{Reconstructed coronal magnetic configuration:
(\textbf{a})~side view, (\textbf{b})~top view. The gray-scale
background shows the magnetogram at the photospheric level. The red
and green field lines represent two arcades. The blue lines
represent the field lines of the excessive magnetic flux that is
rooted in the remote eastern N-polarity sunspot and its environment.
The axes show arc seconds from the center of the region considered.}
  \label{F-extrap}
  \end{figure}

The analysis of the observations and magnetic fields reconstructed
in the corona allowed us to infer the coronal configuration, where
the two events occurred, and their scenario that is considered in
the next section. Figure~\ref{F-extrap} presents the coronal
configuration reconstructed  from an SDO/HMI magnetogram in the
potential-field approximation. Two sets of low loops, which are
rooted in the extended plage region, go into opposite directions
that indicates the presence of a separatrix surface between them.
One set of loops connects the plage region with sunspot Ss1 and the
other connects the plage region with sunspot Ss2. The magnetic flux
coming into the S-polarity sunspots is not compensated by the flux
outgoing from the N-polarity plage region. An excessive portion of
the S-polarity flux is outgoing from the remote N-polarity sunspot
Ss3 and its environment. The magnetic configuration suggests the
possible presence of a coronal null point or even a portion of a
null line where the magnetic-field magnitude $|\textbf{\textit{B}}|$
is zero.

To verify these conjectures, we investigate the spatial distribution
of the coronal magnetic field computed above the active region using
the potential approximation (e.g. \citealp{Uralov2006, Uralov2008};
\citealp{Grechnev2014_I}) and non-linear force-free reconstruction
that is based on the optimization method by \cite{Wheatland2000} as
implemented by \cite{RudenkoMyshyakov2009}. The first stage of our
approach is to analyze the behavior of the null line of the radial
magnetic component $B_r$ at different heights. The presence of a
real null point (which is not an issue of noise or small-scale
uncertainties) is visually indicated by the bifurcation of the $B_r
= 0$ line that occurs as the height changes. The bifurcation appears
as the convergence of the $B_r = 0$ lines followed by their visual
``reconnection'' and subsequent divergence in the orthogonal
direction of the $B_r = 0$ lines that are newly formed. If the
bifurcation occurs in both force-free and potential approximations,
then the null point (or line) is an element of a sufficiently
large-scale magnetic topology. To locate the null point or line
accurately, the behavior of the magnetic-field magnitude
$|\textbf{\textit{B}}|$ in the vicinity of the bifurcation region is
analyzed.

The \url{pot_Br_mod_B.mpg} and \url{nlff_Br_mod_B.mpg} movies in the
Electronic Supplementary Material demonstrate the behaviors of the
$B_r = 0$ line (red) and $|\textbf{\textit{B}}|$ (gray scale and
green contour) in the potential and force-free approximations,
respectively. We are interested in the region of the null point that
is denoted with a circle in the potential approximation and with two
slanted crosses in the force-free approximation. In the latter case,
the only null point exists (lower cross) instead of a null line.
From the null point, a line toward the second cross extends with
$B_r$ nearly zero, but with a weak quasi-transversal magnetic
component. The heights of the null points are noticeably different
in the potential (between 16 and 17\,Mm) and force-free (between 23
and 24\,Mm) approximations.

The configuration presented in Figure~\ref{F-extrap} is similar to
the quasi-circular configuration with a null point that was
considered previously by \cite{Filippov2009}, \cite{Masson2009}, and
\cite{Meshalkina2009}. \cite{Wang2014} analyzed a configuration
where paired M1.9 and C9.2 three-ribbon flares occurred that were
less powerful than our events and had not caused any noticeable CME.
Unlike our approach, \cite{Wang2014} envisioned the presence of the
null line by inference.

\subsection{Scenario}
 \label{S-scenario}

\begin{figure} 
  \centerline{\includegraphics[width=0.8\textwidth]
   {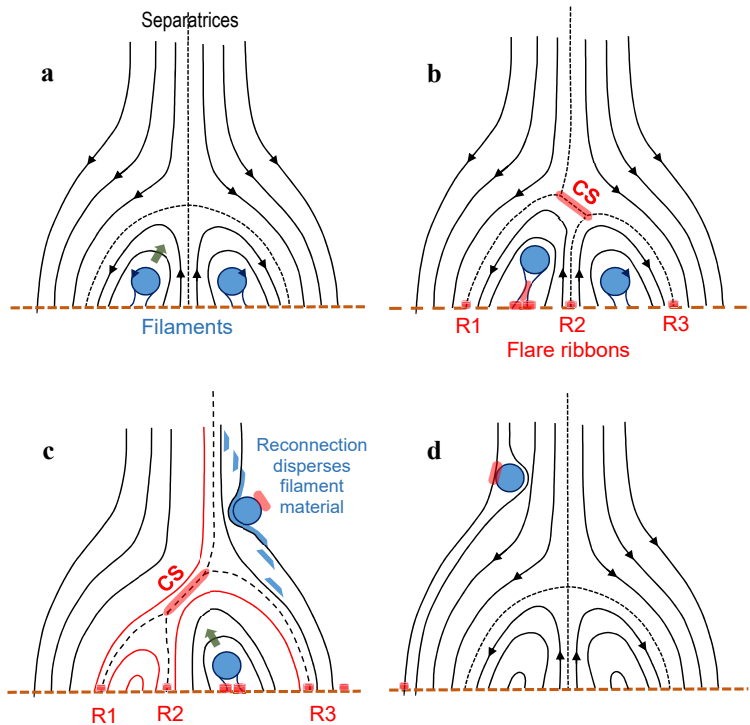}
  }
  \caption{Two-dimensional scheme of the scenario of the two eruptive
events: Side view at the cross-section in the direction indicated in
Figure~\ref{F-mag_config}d by the slanted brown-dashed line. The
solid lines represent magnetic-field lines, the dashed lines
represent separatrices, the blue disks represent cross-sections of
eruptive filaments, and the thick pink bars represent current
sheets. (\textbf{a})~The first filament starts to lift-off.
(\textbf{b})~The rise of the erupting filament results in the
formation of the current sheet (CS). Three flare ribbons R1, R2, and
R3 appear. (\textbf{c})~After the passage of the null, the first
erupting filament (flux-rope progenitor) intensively reconnects in
the outer-right magnetic domain. The current sheet changes
orientation, facilitating the second eruption. (\textbf{d})~The
second filament passes the null and also undergoes reconnection.
After that, relaxation of the configuration to the steady state
begins.}
  \label{F-scenario}
  \end{figure}

Our analysis of observations in different spectral ranges has led to
a scenario of the twin events. For simplicity we consider the 2D
geometry that is justified to some extent by the presence near the
magnetic null point of an extended (about 15\,Mm) linear region
where magnetic field is weak. The scenario is schematically shown in
Figure~\ref{F-scenario} that presents magnetic domains separated
from each other by sepatratrices (broken lines). Each of the inner
domains contains a filament (blue disk). In
Figure~\ref{F-scenario}a, filament 1 in the inner-left domain starts
to lift-off gradually during the initiation phase, while filament 2
in the inner-right domain remains static so far.

Reconnection inside the inner-left domain in
Figure~\ref{F-scenario}b starts transforming filament~1 into a flux
rope. A current instability of this flux-rope progenitor (henceforth
flux rope) triggers its sharp eruption with a motion directed to the
null point. The current sheet (pink bar) forms from the null point
and transfers the magnetic flux into adjacent domains, displacing
the separatrices. Reconnection beneath the erupting structure
reinforces flare energy release (not shown). Accelerated particles
stream down along the field lines, precipitate in dense layers, and
produce HXR emission. Three main flare ribbons R1, R2, and R3
appear.

After the passage of the null in Figure~\ref{F-scenario}c, the first
erupting flux rope enters the outer-right domain where its azimuthal
magnetic component becomes antiparallel to the coronal environment.
The flux rope intensively reconnects with environment and transfers
kinetic energy and magnetic helicity to the reconnected
magnetoplasma. Portions of the filament material are released (blue)
and scatter along the field lines. New field lines (red) appear in
adjacent domains. The current sheet changes orientation that favors
the recovery of their positions. This trend facilitates the second
eruption.

Then, filament~2 erupts and repeats the history of filament~1. After
the passage of the null in Figure~\ref{F-scenario}d, the second flux
rope also undergoes reconnection that results in the dispersal of
its material. Eventually, the configuration relaxes and returns to
the original state.

The features of this scenario are determined by properties of the
coronal configuration. Quasi-circular funnel-like configurations
with magnetic null points exist ubiquitously above small magnetic
islands within the opposite-polarity environment. Nevertheless,
three-ribbon flares are very rare. The particularity of the
configuration in our case was the presence of a coronal magnetic
quasi-null line with almost zero radial component and small
transversal components that made the geometry quasi-two-dimensional.
In the scenario described in this section, the first eruption causes
the second that was also the case in the events addressed by
\cite{Wang2014}.

\section{Manifestations of Accelerated Electrons}
 \label{S-acc_electrons}

\subsection{Hard X-Ray and Microwave Images}
 \label{S-HXR_mw_images}

As shown in Section~\ref{S-microwaves_HXR}, the twin events emitted
conspicuous microwaves, hard X-rays, and $\gamma$-rays that are
produced by accelerated electrons. We compare their manifestations
with EUV images using microwave NoRH \citep{Nakajima1994} and HXR
RHESSI \citep{Lin2002} data. The NoRH enhanced-resolution images at
17\,GHz were synthesized by the \textsf{Fujiki} imaging software.
Note that the spatial resolution of SDO/AIA is one order of
magnitude higher than that of NoRH and two orders of magnitude
higher than that of SRH in its present 48-antenna configuration. For
the imaging in two HXR energy bands, 25\,--\,50\,keV and
50\,--\,100\,keV, we applied the CLEAN method using RHESSI detectors
3 and 8, which operated at that time.

\begin{figure} 
  \centerline{\includegraphics[width=\textwidth]
   {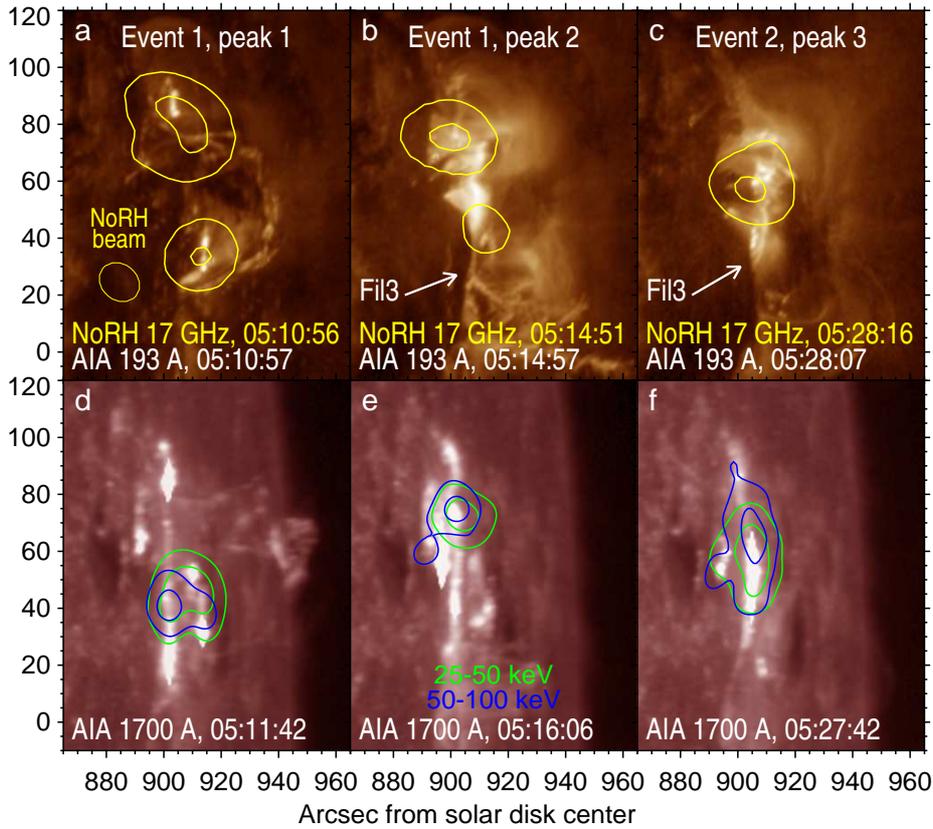}
  }
  \caption{Non-thermal manifestations in microwaves (top) and hard
X-rays (bottom) at three main peaks in comparison with the
structures observed in EUV. (\textbf{a}\,--\,\textbf{c})~Coronal
193\,\AA\ AIA images overlaid with 17\,GHz NoRH images (yellow
contours) at [$0.2, 0.8$] of the maximum values. The thin yellow
contour on panel \textbf{a} represents the NoRH beam at a
half-height level. (\textbf{d}\,--\,\textbf{f})~Flare ribbons in
1700\,\AA\ AIA images along with RHESSI images at 25\,--\,50\,keV
(green contours) and 50\,--\,100\,keV (blue contours) at [$0.3,
0.7$] of the maximum values.}
  \label{F-hxr_mw_images}
  \end{figure}

The AIA 193\,\AA\ images (color background) in Figures
\ref{F-hxr_mw_images}a\,--\,\ref{F-hxr_mw_images}c show the flare
arcades that are partly hidden by filament Fil3. The yellow contours
of the NoRH 17\,GHz images represent non-thermal emissions produced
by accelerated electrons gyrating in the arcade loops. Their
comparison with AIA 193\,\AA\ images is complicated by insufficient
pointing and coalignment accuracy of NoRH images that we enhanced to
about $10^{\prime \prime}$. Nevertheless, it is clear that each of
the microwave sources corresponds to several unresolved arcade
loops, while the source in Figure~\ref{F-hxr_mw_images}c seems to be
a superposition of two arcades.

The AIA 1700\,\AA\ images (color background) in Figures
\ref{F-hxr_mw_images}d\,--\,\ref{F-hxr_mw_images}f show the flare
ribbons at the bases of the coronal arcades. These images are least
distorted by saturation and do not correspond exactly to the times
of the peaks. The HXR sources presented by the blue and green
contours occupy sufficiently bright minor parts of the ribbons. An
extended 50\,--\,100\,keV source in Figure~\ref{F-hxr_mw_images} is
almost as long as the middle ribbon R2. Smaller parts of the
50\,--\,100\,keV sources correspond to ribbons R1 and R2, where
magnetic fields were stronger than those beneath ribbon R2.
Observations of extended ribbon-like structures in hard X-rays are
exceptional (e.g. \citealp{MasudaKosugiHudson2001, Liu2007,
Liu2013}).

\cite{MasudaKosugiHudson2001} pointed out the limitations on the
sensitivity and dynamic range (typically $\approx 10$) in the HXR
imaging that Fourier-synthesis telescopes provide because of a
limited coverage of the $(u,v)$-plane. \cite{Krucker2014} confirmed
this conclusion for RHESSI; sources weaker than about 10\,\% of the
brightest in an HXR image are not detectable. Thus, the simplicity
and compactness of HXR sources that are usually observed result from
instrumental limitations and do not reflect properties of
non-thermal processes. Conversely, rare observations of ribbon-like
HXR sources indicate some atypical conditions such as the uniformity
of the magnetic-field strength or the uniformity of electron
acceleration along the arcade \citep{Liu2007}.

Microwave and HXR images highlight different parts of the flare
configuration, where accelerated electrons were present, but reveal
them incompletely. The NoRH images at 17\,GHz present unresolved
sets of coronal loops, where magnetic field is sufficiently strong.
Weaker-field regions are fainter at 17\,GHz. With the power-law
indices of the HXR spectra [$\gamma$] found in the next section, the
emission at optically thin frequencies is expected to be highly
dependent on the magnetic-field strength [$B$] as $B^{0.9\delta -
0.22} \approx B^4$; $\delta = \gamma + 1.5$ \citep{DulkMarsh1982,
White2011}. Microwaves and hard X-rays dominate at different sides
of an asymmetric magnetic configuration; the magnetic mirroring
impedes electron precipitation (and hence the HXR emission) at the
stronger-field side, where the optically thin gyrosynchrotron
emission is stronger \citep{Kundu1995}.

These circumstances indicate that manifestations of accelerated
electrons correspond to the structures observed in the EUV, but they
are revealed incompletely because of instrumental limitations of
RHESSI and NoRH. This result corresponds to the conclusions drawn
previously by \cite{Zimovets2013} and \cite{Grechnev2017}.

\subsection{Hard X-Ray and Microwave Spectra}
 \label{S-mw_spectra}

The HXR spectra for the three temporal intervals corresponding to
the main peaks were computed from RHESSI data using detectors 3 and
8 by means of the \textsf{OSPEX} package. The spectra in the
3\,--\,250\,keV range have typical shapes that are consistent with a
thermal core and non-thermal tail. The spectra were fitted with a
thermal component at low energies and single power-law component at
higher energies. Table~\ref{T-spectra} presents the
accumulation-time intervals and parameters found for the spectra,
the power-law index [$\gamma$] and the normalization constant
[$A_0$] at a fiducial energy of 50\,keV. The relations between the
HXR spectra of the three peaks confirm the qualitative conclusions
drawn in Section~\ref{S-microwaves_HXR} from the RHESSI temporal
profiles presented in Figure~\ref{F-time_prof}.

\begin{table} 
 \caption{Parameters of the HXR and microwave spectra.}
 \label{T-spectra}
 \begin{tabular}{lccccccc}

 \hline

 &  \multicolumn{3}{c}{Hard X-rays} &  \multicolumn{4}{c}{Microwaves}  \\

 &  \multicolumn{3}{c}{\hrulefill} &  \multicolumn{4}{c}{\hrulefill}  \\

 & Accumulation & $A_0$\tabnote{[photons s$^{-1}$ cm$^{-2}$ keV$^{-1}$]}  &  $\gamma$  & $N_{\mathrm{r}}$\tabnote{[$10^7$\,electrons\,cm$^{-3}$]} & $\overline{|B|}$ & $\nu_\mathrm{peak}$ & $S_{\max}$ \\

 & time [UTC] &  at 50\,keV &  & $\geq 10$\,keV  & [G] & [GHz] & [sfu] \\

 \hline

Peak~1 & 05:09:40--05:11:43 & 1.08 \  &  3.33  & 7.42 & 480 & \ 6.1 & 370 \\

Peak~2 & 05:14:40--05:15:41 & 0.375 &  3.03  & 3.23 & 630 & 12.4 & 390 \\

Peak~3 & 05:27:12--05:29:00 & 0.748 &  3.16  & 3.44 & 350 & \ 7.4 & 720 \\

 \hline

 \end{tabular}
 \end{table}

With the parameters found for the HXR spectra, it is possible to
model the microwave spectra using the magnetogram actually observed.
The most advanced modeling tool, of which we are aware, is the
\textsf{GX Simulator} developed by \cite{Nita2015}. In particular,
this software allows one to model the gyrosynchrotron (GS) spectrum
of an inhomogeneous source with various electron distributions based
on a real magnetogram. However, the usage of the \textsf{GX
Simulator} with a considerable number of loops seems to be
difficult.

We invoke a different approach following \cite{Grechnev2017} who
used a multi-loop model. The model contains a considerable number of
pairs of homogeneous cubic GS sources in the legs of
microwave-emitting loops rooted in the ribbons, each with a
different magnetic-field strength and volume. In our flare, one of
the legs of each loop is rooted in the middle ribbon. The total flux
is the sum of the fluxes emitted by all of the loops, which are
considered not to overlap with each other; thus, the number of the
loops is about the ribbon length-to-width ratio (13 in our case).
The number [$m$] of paired sources is chosen so that the sum of
[$2m$] sources provides a smooth spectrum with single maximum and
the modeled spectrum acceptably matches the observed fluxes measured
at different frequencies. The model uses a simplified analytical
description of the GS emission presented by \cite{DulkMarsh1982}.
Using the radial magnetic-field distribution on the photosphere and
the balance of magnetic fluxes in the conjugate legs of the loops,
parameters of each source are estimated. Then, the magnetic-field
strengths are corrected to the coronal values using a constant
scaling factor that is estimated by referring to
\cite{LeeNitaGary2009}. The model underestimates the low-frequency
part of the spectrum that we disregard.

From the parameters of the HXR spectra for the three peaks that are
listed in the middle part of Table~\ref{T-spectra}, we estimated the
number $N_{\mathrm{r}}$ of microwave-emitting electrons per unit
volume in the distribution above the fiducial electron energy
$E_{\mathrm{r}} = 10$\,keV \citep{DulkMarsh1982, White2011}. The
average magnetic-field strength in the coronal microwave sources
$\overline{|B|}$, the peak frequency $\nu_\mathrm{peak}$ of the GS
spectrum, and the maximum flux $S_{\max}$ at this frequency found in
the modeling for each of the three peaks are listed in the right
part of Table~\ref{T-spectra}. The magnetic-field strengths
estimated in the modeling do not contradict the values found from
the extrapolation in the potential-field approximation.

\begin{figure} 
  \centerline{\includegraphics[width=0.65\textwidth]
   {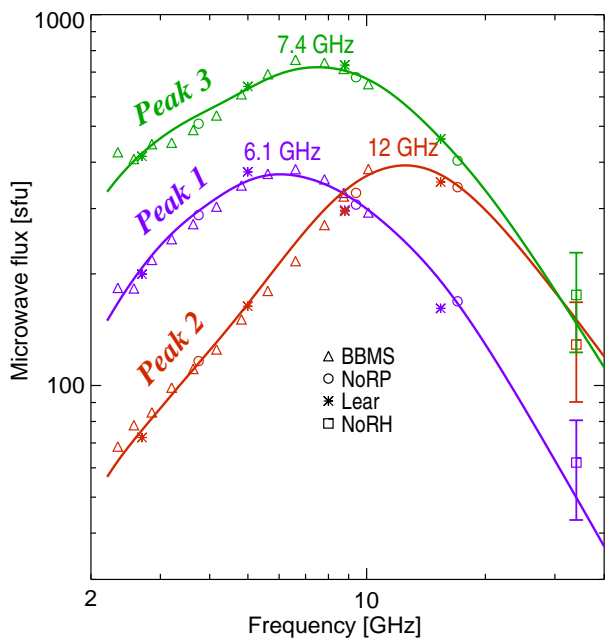}
  }
  \caption{Gyrosynchrotron spectra modeled for the three main peaks
(color curves). The symbols represent the actual measurements (same
as in Figure~\ref{F-mw_spectra}). }
  \label{F-model_spectra}
  \end{figure}

The spectra of the three peaks obtained in the modeling of 13 pairs
of microwave sources are presented in Figure~\ref{F-model_spectra}
along with the observed values (similar to
Figure~\ref{F-mw_spectra}). The modeling acceptably reproduces the
turnover part of the actual spectra and their adjacent branches
between 2 and 17\,GHz. The modeled spectra are underestimated at
2\,GHz and lower frequencies (not shown). It is only possible to
state that the modeled spectra do not contradict the fluxes
estimated at 34\,GHz, because their uncertainties are large. A
formal increase in the number of sources does not bring the modeled
spectra closer to the observations.

The modeling confirms that microwaves were emitted in coronal loops
by the same electron populations, whose precipitation produced hard
X-rays in both flares, although the spatial separation between the
HXR and microwave sources was considerable. The low-frequency
flattening of the microwave spectra was caused by the asymmetry of
the magnetic configuration. The challenging shift of the spectral
turnover to the left-up from peak~2 to peak~3 was a result of an
increased asymmetry in the second flare. The modeling also confirms
that apparently simple microwave sources observed by NoRH at 17\,GHz
actually represented two flare arcades. Manifestations of
accelerated electrons in microwaves correspond to the structures
observed in the EUV indeed.

\section{Summary and Concluding Remarks}
 \label{S-conclusion}

Having got a pointing from SRH at two major flares, which occurred
within half an hour in the same active region and exhibited
different emission spectra, we analyzed the twin events using EUV,
hard X-ray, and microwave data along with magnetograms. Both events
were eruptive flares, each of which had an unusual three-ribbon
configuration. The events led to a CME, whose structure indicated
its origin due to the twin eruptions.

The magnetic configuration, where the events occurred, was
characterized by a considerable excess of the S-polarity magnetic
flux over the N-polarity flux in an extended plage region. The
excessive S-polarity flux high above the plage region was
concentrated within a tube-like outer spine that was rooted in a
remote N-polarity sunspot. The coronal configuration just above the
plage region had a shape of an inverted funnel that contained a null
point in the waist. Because of the extended geometry, the funnel and
the null-point region were stretched along the solar surface
parallel to the plage region.

Each of the two magnetic domains inside the funnel contained a
filament before the events. The filaments erupted one after another
within half an hour. A scenario was inferred from multi-spectral
observations that combines the twin events in terms of
null-point-associated successive eruptions. Each rising filament
moved to the null-field region that inevitably resulted in partial
reconnection between the erupting structure and static coronal
environment. Two successive flares developed in this way that
basically resembled circular-ribbon flares with the following
modifications: The central brightening extended into the middle
ribbon located in the N-polarity plage region and the circular
ribbon transformed into two ribbons located in S-polarity magnetic
fields on both sides of the middle ribbon. Thus, a three-ribbon
flare configuration appeared. In this scenario, the first filament
eruption facilitates the second.

The two flares produced considerable microwave, hard X-ray, and
$\gamma$-ray bursts that are detectable at energies exceeding
800\,keV. Microwave and hard X-ray images highlighted different
parts of the flare configuration, where accelerated electrons were
present. Hard X-ray sources occupied minor parts of the ribbons that
were sufficiently bright in the EUV. An extended hard X-ray source
in the second flare was almost as long as the middle ribbon. Two
microwave sources that resembled two footpoints of a single loop
represented in fact two different arcades that is confirmed by the
modeling of microwave spectra. Manifestations of accelerated
electrons in hard X-rays and microwaves corresponded to the
structures observed in the EUV, but they were revealed incompletely
because of instrumental limitations of RHESSI and NoRH.

The results indicate that the spatial resolution achievable in
microwave observations, which are currently available, may be
insufficient to discern flaring structures. On the other hand, the
dynamic range of hard X-ray imagers may be insufficient to reveal
the structures of hard X-ray sources perfectly. Hence,
higher-resolution images with a sufficient dynamic range obtained in
different spectral domains should be invoked for a correct
interpretation of non-thermal flare sources.

\begin{acks}
We thank L.K.~Kashapova for useful discussions. We appreciate our
colleagues from the Radio Astrophysical Department and the Radio
Astrophysical Observatory in Badary. This study was funded by the
Russian Science Foundation under grant 18-12-00172. The development
of the methods used in Sections \ref{S-microwaves_HXR} and
\ref{S-acc_electrons} was supported by the Program of Basic Research
of the RAS Presidium No.~28. The SRH and BBMS data were obtained
using the Unique Research Facility Siberian Solar Radio Telescope
(\url{ckp-rf.ru/usu/73606}).

We thank the NASA/SDO and the AIA and HMI science teams, the teams
operating RHESSI, the Nobeyama solar facilities, and the US AF RSTN
network for the data used here. We thank the International
Consortium for the continued operation of Nobeyama Radioheliograph
until shut down on 31 March 2020. We are grateful to the team
maintaining the CME Catalog at the CDAW Data Center by NASA and the
Catholic University of America in cooperation with the Naval
Research Laboratory.
\end{acks}

\section*{Disclosure of Potential Conflicts of Interest} The authors
declare that they have no conflicts of interest.

\bibliographystyle{spr-mp-sola}

\bibliography{three_ribbon_flares}

\end{article}

\end{document}